\documentclass[preprint,showpacs,amsmath,amssymb,byrevtex,endfloats*,citeautoscript,altaffilletter,a4paper]{revtex4}

\usepackage{graphicx}
\usepackage{dcolumn}
\usepackage{bm}
\usepackage{helvet} 

\begin{document}


\title{Characterization of chaotic dynamics in the vocalization of  {\em Cervus Elaphus Corsicanus}}

\author{Angelo Facchini}
\email{a.facchini@unisi.it}
 \affiliation{Department of Chemical and Biosystems Sciences,
University of Siena\\
  Via della Diana 2/A 53100 Siena, Italy.}

\author{Simone Bastianoni}
\email{bastianoni@unisi.it}
 \affiliation{Department of Chemical and Biosystems Sciences,
University of Siena\\
  Via della Diana 2/A 53100 Siena, Italy.}

  \author{Nadia Marchettini}
  \email{marchettini@unisi.it}
 \affiliation{Department of Chemical and Biosystems Sciences,
University of Siena\\
  Via della Diana 2/A 53100 Siena, Italy.}

\author{Mauro Rustici}%
\altaffiliation{Corresponding Author}
\email{rustici@uniss.it}
\affiliation{Department of Chemistry,  University of Sassari, Via Vienna 2-I 07100 Sassari, Italy.}%


\date{\today}

\begin{abstract}
Chaos, oscillations, instabilities, intermittency  represent only
some nonlinear examples apparent in natural world. These phenomena
appear in any field of study, and advances in complex and
nonlinear dynamic techniques bring about opportunities to better
understand animal signals.

In this work we suggest an analysis method based on the
characterization of the vocal fold dynamics by means of the
nonlinear time series analysis, and by the computations of the
parameters typical of  chaotic oscillations: Attractor
reconstruction, Spectrum of Lyapunov Exponents and Maximum
Lyapunov Exponent was used to reconstruct the dynamic of the vocal
folds. Identifying a sort of of vocal fingerprint can be useful in
biodiversity monitoring and understanding the health status of a
given animal.

This method was applied to the vocalization of the {\em Cervus
Elaphus Corsicanus}, the Sardinian Red Deer.
\end{abstract}

\maketitle

\section{Introduction}

The physical and physiological mechanism of sound production are
important to understanding mammal vocalization which ranges from
periodic vocal fold vibrations to completely aperiodic vibration
and atonal noise. Between these two extremes, a large amount of
phenomena have been observed and reported \cite{m2,m4}:
biphonation, cycles, subharmonic and chaotic behavior. These
behaviors can be predicted by theoretical models. For example, the
two mass model (the most accepted for mammal apparatus of
phonation) can exhibit irregular oscillations
\cite{2masse,jasa110-7}.
 The apparatus of phonation can be investigated through the
characterization of the animal vocalization, where vocal
nonlinearity can be used. According to Tokuda\cite{toku} the
nonlinear analysis of human speech signal has been carried out
extensively, while nonlinear characteristics for animal voice
signals have not yet been investigated.

Using the methods of nonlinear time series analysis we wished to
understand the mechanics of the vocal folds starting from the
vocalization time series. The characterization  of the vocal
signal as a chaotic time series can give important information on
the health status of the animal, since the oscillation modes are
related to the status of the throat tissues and to the strength of
the animal. Furthermore  tissues shapes of vocal apparatus are
different among the animals and  the characterization of several
chaotic signals can be used in the monitoring of biodiversity.

The last remaining populations of a sub-specie of the red deer:
the Sardinian deer ({\em Cervus Elaphus Corsicanus}) are found in
the well preserved evergreen forest of {\em Monte Arcosu} in
Sardinia (a protected area owned  by WWF Italy). The {\em Cervus
Elaphus} is the largest and most phylogenetically advanced species
of Cervus. Head and body length is 1.65-2.65m, tail length is
0.11-0.27m, height at the shoulder is 0.75-0.15m, and weight is
75-340 kg. The largest and strongest male generally has the
largest harem. In order to maintain this position of superiority
he must constantly keep the distance with rival males by bellowing
out, and chasing off potential rivals who come near his females.
After vocalizing, the largest remaining males size each other up,
and if antler and body size are comparable, they battle for the
females. Their antlers lock and each male attempts to forcefully
push the other away. The strongest and most powerful male wins and
secures a harem (group) of females for mating. In this work  an
extensive characterization of the vocalization of  {\em Cervus
Elaphus Corsicanus}  is presented by means of Lyapounov exponents
of the chaotic oscillation evidence  of registered sounds.

\section{Material}
\begin{figure}
\includegraphics[height=12cm, keepaspectratio=true]{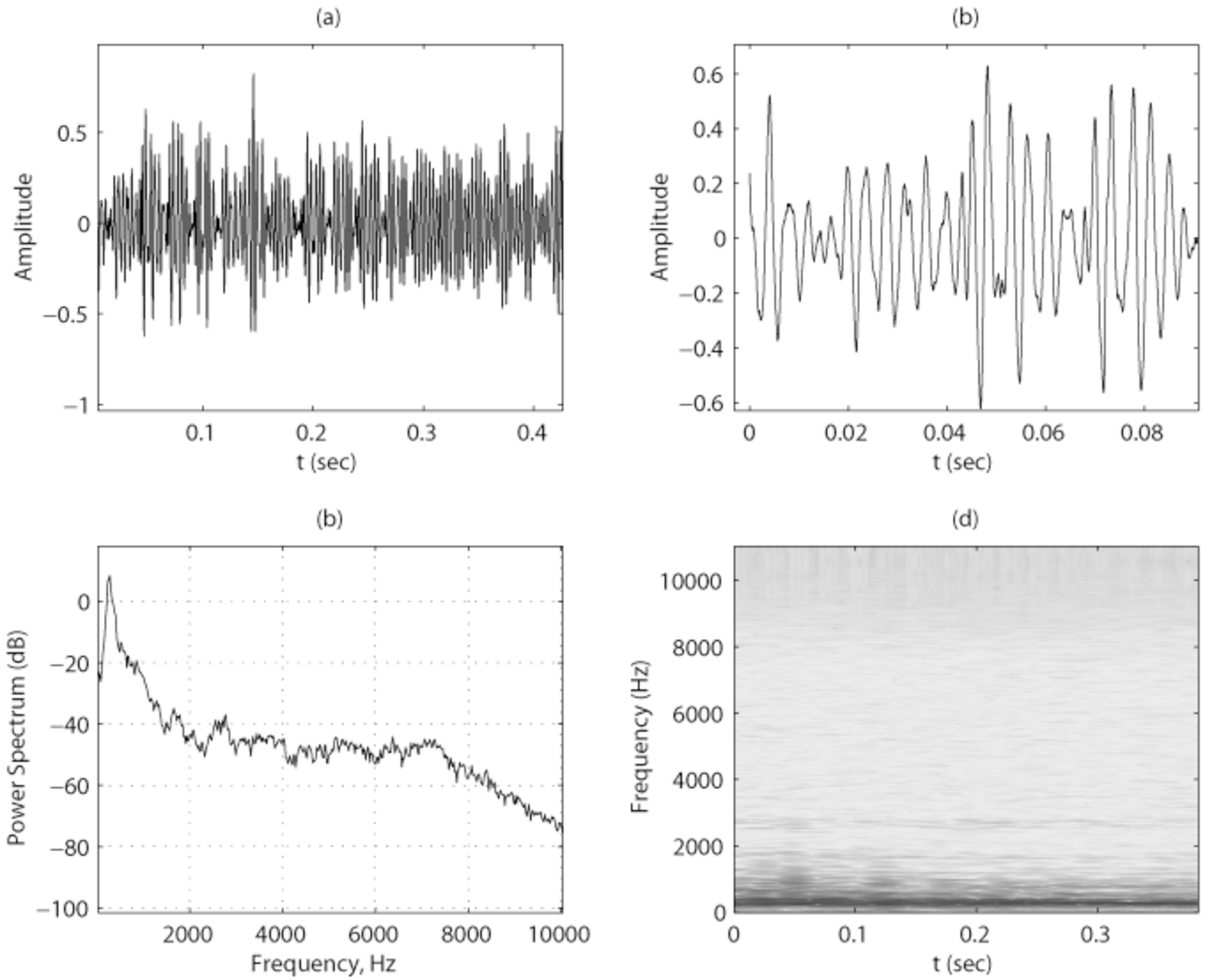}
\caption{(a) The analyzed signal. (b) A portion of the signal
showing the irregular nature of the vocalization. (c) The power
spectrum shows the typical Spectral contents of an irregular
signal: a broadband and a continuous one. (d) The spectrogram of
the signal showing a fundamental frequency of 70Hz and other
frequencies until 2000Hz} {\label{fig1}}
\end{figure}

A number of different signals corresponding to different sound
emissions were considered. Only clear and low noise sound
emissions have been analyzed, in order to focus exclusively on
meaningful vocalizations, and to avoid  spurious effect. The
vocalizations were recorded from  adult males in their natural
environment and digitized with a sampling frequency of 22050 Hz.
Fig.\ref{fig1}(b) shows a small portion of the analyzed signal and
in Fig.\ref{fig1}(d), the spectrogram (512 points FFT) of the
signal is shown.

Discrete Fourier Transform (Fig.\ref{fig1}(c))  was used to
perform a preliminary spectral analysis on vocalization units. The
presence of regions with  high density of unresolved frequencies
is a necessary, even if not sufficient, condition for the
occurrence of chaotic dynamical regimes \cite{ott93}. Non-linear
dynamics analysis were, therefore, was limited to signal units
characterized by broad-band features in the frequency domain.
Results reported in the present work refer to a single signal
0.420s long.  The time series examined consists of a 9455 points
sampled at 22050Hz.

\section{Computational methods}
The analysis of the time series was performed using the software
package TISEAN\footnote{The TISEAN software package is publicy
available at
http://www.mpipks-dresden.mpg.de/$\sim$tisean/TISEAN$\_2.1$/index.html.}
(TIme SEries ANalysis) \cite{Kantz97}, valued as the most well
known and robust algorithm set for nonlinear time series analysis.
Typical steps are attractor reconstruction from time series and
the characterization of the chaotic dynamic by means  of Lyapunov
exponents and maximum Lyapunov exponent (MLE).

\subsection{Attractor reconstruction}
The attractor of underlying dynamics has been reconstructed in
phase space by applying the time delay vector method
\cite{ott93,aba96}.

Starting from a time series $s(t)=[s_1,\dots,s_N]$ the system
dynamic can be reconstructed using the delay theorem by Takens and
Ma\~{n}e. The reconstructed trajectory $\mathbf{X}$  can be
expressed as a  matrix where each row is a phase space vector:

\begin{equation}
\mathbf{X}=[X_{1},X_{2},\dots,X_{M}]^T
\end{equation}
where $X_{i}=[s_{i},s_{i+T},\dots,s_{i-(D_{E}-1)T}]$ and
$M=N-(D_{E}-1)T$.

The matrix is characterized by two key parameters: The {\em
Embedding Dimension} $D_{E}$ and the {\em Delay Time} $T$. The
embedding dimension is the minimum dimension at which the
reconstructed attractor can be considered completely unfolded and
there is no overlapping in the reconstructed trajectories. If the
chosen dimension is lower than $D_{E}$ the attractor is not
completely unfolded and the underlying dynamics cannot be
investigated. Higher dimension was not used due to  the increase
in computational effort.

The algorithm used for the computation of $D_{E}$ is the method of
{\em False Nearest Neighbors}\cite{aba93}. A false neighbor is a
point of trajectory intersection in a poorly reconstructed
attractor. As the dimension increases, the attractor is  unfolded
 with greater fidelity, and the number of false neighbors decreases to zero. The first
dimension with no overlapping points is $D_E$.

The delay time $T$ represents a measure of correlation existing
between two consecutive components of $D_{E}$-dimensional vectors
used in the trajectory reconstruction. Following a commonly
applied methodology, the time delay $T$ is chosen in
correspondence to the first minimum of the average mutual
information function \cite{fra86}.

\subsection{Lyapunov exponents}
Chaotic systems display a sensitive dependence on initial
conditions. Such a property deeply affects the time evolution of
trajectories starting from infinitesimally close initial
conditions, and Lyapunov exponents are a measure of this
dependence. These characteristic exponents give a coordinate
independent measure of the local stability properties of a
trajectory. If the trajectory evolves in a $N$-dimensional state
space there are $N$ exponents arranged in decreasing order,
referred to as the {\em Spectrum of Lyapunov Exponents (SLE)}:
\begin{equation}
\lambda_1 \ge \lambda_2 \ge \dots \ge \lambda_{n}
\end{equation}
Conceptually these exponents are a generalizations of eigenvalues
used to characterize different types of equilibrium points.

A trajectory is chaotic if there is at least one positive
exponent,  the value of this exponent, said the {\em Maximum
Lyapunov Exponent (MLE)} gives a measure of the divergence rate of
infinitesimally close trajectories and of the unpredictability of
the system and gives a good characterization of the underlying
dynamics.

Starting from the reconstructed attractor $\mathbf{X}$, it is
possible to compute with the method of Sano and
Sawada\cite{Greene87,Sano85} the SLE consisting of exactly
$n=D_{E}$ exponents. This method is a qualitative one, and in
presence of a positive exponents, $\lambda_1$, a more accurate
method is necessary for the computation.

The method of Rosenstein-Kantz\cite{rose93,kantz94} is used to
compute the MLE from the time series. This method measures in the
reconstructed attractor  the average divergence of two close
trajectories in the time $d_{j}(i)$. This can be expressed as:
\begin{equation}
d_{j}(i)=C_{j}e^{\lambda_{1}(i\Delta t)}
\end{equation}
where $C_{j}$ is the initial separation. By taking the logarithm
of both sides we obtain:
\begin{equation}
\ln d_{j}(i)=\ln C_{j} +\lambda_{1}(i\Delta t)
\end{equation}
This is a set of approximately parallel lines (for $j=1,2,\dots,
M$) each with a slope roughly proportional to $\lambda_{1}$. The
MLE is easily calculated using a least-squares fit to the average
line defined by
\begin{equation}
y(i)=\frac{1}{\Delta t} \langle \ln d_{j}(i)\rangle
\end{equation}
where $\langle \cdot \rangle$ denotes the average over all values
of $j$. Figure \ref{fig2}(d) shows a typical plot of $\langle \ln
d_{j}(i)\rangle$: after a short transition there is a linear
region that is used to extract the MLE.

\section{Results and Discussion}
The signal considered was characterized by highly complex patterns
in which different transients with both periodic and apparently
aperiodic features were identified. The apparently random behavior
of the numerical series, easily detectable with a simple visual
inspection of the sound pattern, was confirmed by the power
spectrum and  spectrogram. Three different regions were put into
evidence: at low frequencies, between 0 and 70 Hz, a first
distribution of unresolved peaks is present, a sharp peak is also
present at 450 Hz, while a broad band of frequencies, ranging
between 850 and 1500 Hz, is easily detectable.

\begin{figure}
\begin{center}
\includegraphics[height=12cm, keepaspectratio=true]{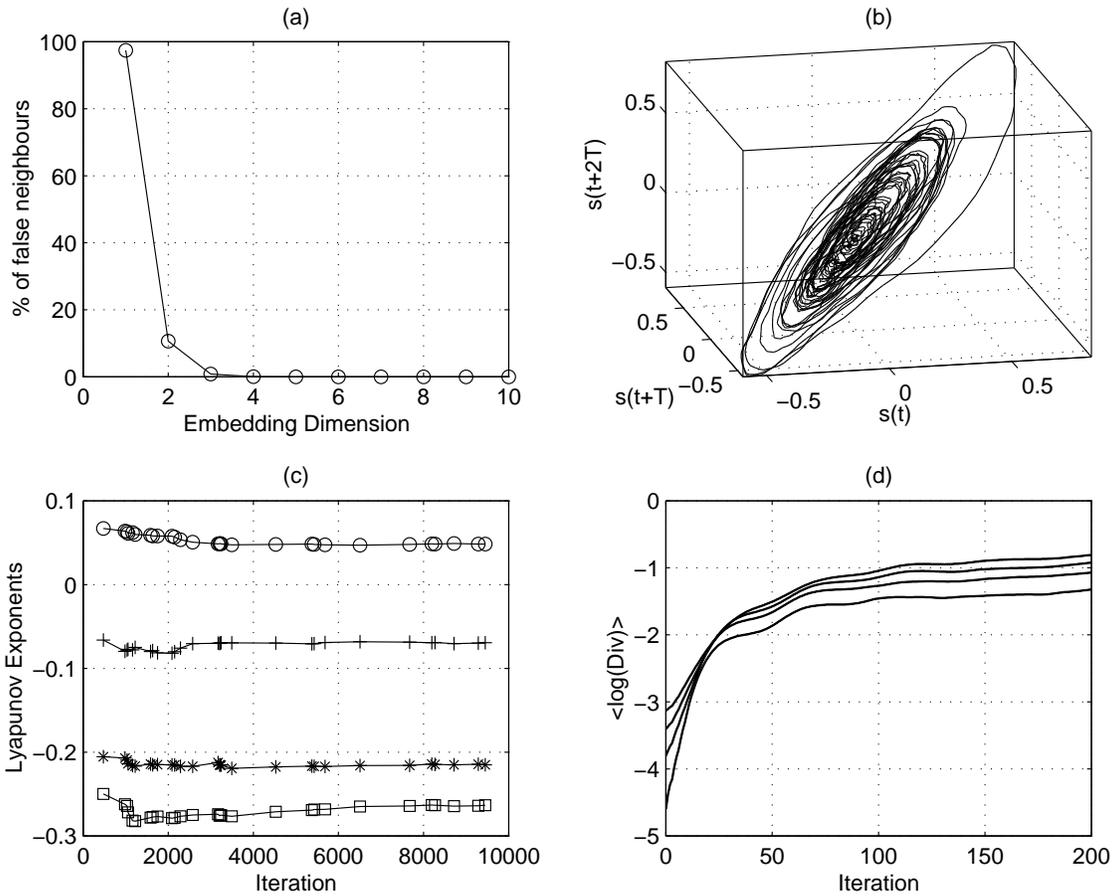}
\caption{(a) Computation of the embedding dimension by the False
Nearest Method applied to the time series. The fraction of false
neighbors decrease to zero at a reconstruction dimension $D_E=4$.
(b) The attractor reconstructed by the method of delays: This
highly structured trajectory indicates that the signal is chaotic
and that the irregular motion is not a noisy one. (c) The Spectrum
of Lyapunov Exponent showing the presence of a positive Lyapunov
exponent and three negative exponents. (d) Computation of the
Maximum Lyapunov Exponent by the Rosenstein-Kantz algorithm. The
value of $\lambda_{1}$ is obtained by a linear regression of the
curves in the zone between 0 and 20 iterations. The value
$\lambda_{1}=0.48$ was found.
 }
{\label{fig2}}
\end{center}
\end{figure}

The chaotic characterization was performed calculating the
embedding dimension $D_{E}$ by the false nearest method and in
Fig.\ref{fig2}(a) the result of the computation is shown. The
figure reports the fraction of false neighbors with respect to the
embedding dimension and a value of $D_{E}=4$ was found. The delay
Time was considered as the first minimum of the mutual information
function, and the value $T=8$ was found.

Starting from the time series the attractor was reconstructed
using the delay method, and in Fig.\ref{fig2}(b) a three
dimensional projection of the attractor is shown. The structure of
the attractors, related to the chaotic oscillation of the vocal
folds, demonstrated that the irregular behavior observed in the
time series was not due to noise.

In order to completely characterize the chaotic nature of the
vocalization, the Spectrum of Lyapunov Exponents and the Maximum
Lyapunov Exponent $\lambda_{1}$ were evaluated. In
Fig.\ref{fig2}(c) values of the four exponents are reported and
the presence of a positive exponent was detected. The accurate
value of the MLE was computed by the Rosenstein-Kantz method and a
value of $\lambda_{1}=0.48$ was found by a linear regression of
the curves in the region between 0 and 20 iterations.

The Kaplan-Yorke fractal dimension $D_L$ of the attractor
\cite{kap79}, equal to $D_L = 2.58$, confirms the high dimensional
fractal qualities of the strange attractor.

\section{Concluding remarks}
The analysis method proposed in this letter was applied to the
vocalization of an adult male of {\em Cervus elaphus corsicanus}
and put in evidence the chaotic behavior of the irregolar
oscillations  in the signal considered. A full characterization by
means of attractor reconstruction,Spectrum of Lyapunov Exponents,
and Maximum Lyapunov Exponent was performed. A positive value of
MLE was found. Future work aimed at identifying different
individuals through the discussed parameters, will consist in the
analysis of other vocalizations looking for a {\em vocal
fingerprint} that may be useful in biodiversity monitoring.


\begin{acknowledgments}
The authors are thankful to Dr. Carlo Murgia (Director of Oasi di
monte Arcosu Sardegna) for providing the vocalizations.
\end{acknowledgments}

%

\begin{table}
\caption{\label{tab} Results of the analysis performed on the
vocalization signal. A positive Lyapunov exponent and the value of
Kaplan-Yorke dimension indicates the chaotic nature of the
signal.}
\begin{ruledtabular}
\begin{tabular}{lc}
 Parameter & Value \\
\hline
Delay Time $\mathbf{T}$ & 8 \\
Embedding Dimension $\mathbf{D_{E}}$ & 4 \\
Maximum Lyapunov Exponent $\mathbf{\lambda_{1}}$ & 0.48  \\
Kaplan-Yorke Dimension $\mathbf{D_{L}}$& 2.58 \\
\end{tabular}
\end{ruledtabular}
\end{table}

\bibliography{letterJ}

\begin{thebibliography}{10}

\bibitem{m2}
I.Wilden, H.Herzel, G.Peters, and G.Tembrock, ``Subharmonics, biphonation and
  deterministic chaos in mammal vocalization,'' Bioacoustics,171--196 (1998).
\vspace{.125in}
\bibitem{m4}
I. Steinecke and H. Herzel, ``Bifurcations in an asymmetric vocal fold model,''
  J. Acoustical Soc. of America {\bf 97}(3), 1874--1884 (1995).
\vspace{.125in}
\bibitem{2masse}
H.Herzel, D.Berry, I.Titze, and I.Steincke, ``Nonlinear dynamics of the voice:
  Signal analysis and biomechanical modeling,'' CHAOS {\bf 5}(1), 30--34
  (1995).
\vspace{.125in}
\bibitem{jasa110-7}
J.J.Jiang, Y.Zhang, and J.Stern, ``Modeling of chaotic vibrations in symmetric
  vocal folds,'' J. Acoust. Soc. Am. {\bf 110}(4), 2120--2128 (2001).
\vspace{.125in}
\bibitem{toku}
I.Tokuda, T.Reide, J.Neubauer, M.J.Owren, and H.Herzel, ``Nonlinear analysis of
  irregular animal vocalizations,'' J. Acoust. Soc. Am {\bf 111}(6), 2908--2919
  (2002).
\vspace{.125in}
\bibitem{ott93}
E.Ott, {\sl Chaos in dynamical systems} (Cambridge University Press, UK, 1993).
\vspace{.125in}
\bibitem{Kantz97}
H.Kantz and T.Schreiber, {\sl Nonlinear Time Series Analysis} (Cambridge
  University Press, UK, 1997).
\vspace{.125in}
\bibitem{aba96}
H.D.I.Abarbanel, {\sl Analysis of observed chaotic data} (Springer-Verlag,
  ADDRESS, 1996).
\vspace{.125in}
\bibitem{aba93}
H.D.I.Abarbanel and M.B.Kennel, ``Local false nearest neighbours and dynamical
  dimensions from observed chaotic data,'' Phys. Rev. E {\bf 47}, 3057 (1993).
\vspace{.125in}
\bibitem{fra86}
A.M.Fraser and H.L.Swinney, ``Independent coordinates for strange attractors
  from mutual information,'' Phys. Rev. A {\bf 33}, 1134 (1986).
\vspace{.125in}
\bibitem{Greene87}
J.M.Greene and J.S.Kim, ``The calculation of lyapunov spectra,'' Physica D {\bf
  24}, 213--225 (1987).
\vspace{.125in}
\bibitem{Sano85}
M.Sano and Y.Sawada, ``Measurement of the lyapunov spectrum from a chaotic time
  series,'' Phys. Rev. Lett. {\bf 55}, 1082 (1985).
\vspace{.125in}
\bibitem{rose93}
M.T.Rosenstein, J.J.Collins, and C. Luca, ``A practical method for calculating
  largest lyapunov exponent from small data set,'' Physica D {\bf 65}, 117
  (1993).
\vspace{.125in}
\bibitem{kantz94}
H. Kantz, ``A robust method to estimate the maximal lyapunov exponent of a time
  series,'' Phys. Lett. A {\bf 185}, 77 (1994).
\vspace{.125in}
\bibitem{kap79}
J. Kaplan and J.~A. Yorke, {\sl Chaotic behaviour of multidimensional
  difference equation}, Vol.~730 of {\sl lecture notes in mathematics}
  (Springer Verlag, Berlin, 1979).
\vspace{.125in}
\end{thebibliography}
\bibliographystyle{jasasty}


\end{document}